\documentclass{elsart}

\usepackage{amsmath,amsfonts,amssymb}
\usepackage{mathrsfs,bbm}                               
\usepackage{graphicx,subfigure,pst-all%
,pstricks-add,psfrag}					

\usepackage[sort&compress]{natbib}			
\bibpunct{[}{]}{,}{n}{}{;}				

	\newcommand{\E}[1]{\ensuremath{\mathbb{E}\!\left[#1\right]}}

	\newcommand{\abs}[1]{\ensuremath{\left|#1\right|}}

	\DeclareMathOperator{\sgn}{sgn}

	\renewcommand{\Re}{\ensuremath{\mathbb{R}}}
	\newcommand{\N}{\ensuremath{\mathbb N}}

	\DeclareMathAlphabet{\mathpzc}{OT1}{pzc}{m}{it}
\begin{document}
\begin{frontmatter}
\title{Wavelet entropy and fractional Brownian motion time series}
\author[pucv]{D. G. P\'erez\corauthref{cor}},
\ead{dario.perez@ucv.cl}
\author[ciop]{L. Zunino},
\corauth[cor]{Corresponding author.}
\ead{lucianoz@ciop.unlp.edu.ar}
\author[ciop,fis]{M. Garavaglia}
\ead{garavagliam@ciop.unlp.edu.ar}
and
\author[uba]{O. A.  Rosso}
\ead{oarosso@fibertel.com.ar}

\address[pucv]{Instituto de F\'isica, Pontificia Universidad Cat\'olica de Valpara\'iso (PUCV), 23-40025 Valpara\'iso, Chile.}

\address[ciop]{Centro de Investigaciones \'Opticas (CIOp), CC. 124 Correo Central,1900 La Plata, Argentina.}

\address[fis]{Departamento de F\'{\i}sica, Facultad de Ciencias Exactas, Universidad Nacional de La Plata (UNLP), 1900 La Plata, Argentina.}

\address[uba]{Instituto de C\'alculo, Facultad de Ciencias Exactas y Naturales, Universidad de Buenos Aires (UBA), Pabell\'on II, Ciudad Universitaria, 1428 Ciudad de Buenos Aires, Argentina. }


\begin{abstract} 
We study the functional link between the Hurst parameter and the Normalized Total Wavelet Entropy when analyzing fractional Brownian motion (fBm) time series---these series are synthetically generated. Both quantifiers are mainly used to identify fractional Brownian motion processes (Fractals 12 (2004) 223). The aim of this work is understand the differences in the information obtained from them, if any.
\end{abstract}

\end{frontmatter}

\section{Introduction}\label{sec:intro}

When studying the laser beam propagation through a laboratory-generated turbulence~\cite{paper:zunino2004} we have introduced two quantifiers: the \textit{Hurst parameter}, $H$, and the \textit{Normalized Total Wavelet Entropy} (NTWS), $S_\text{WT}$. The former quantifier was introduced to test how good the family of \textit{fractional Brownian motion}~\cite{paper:mandelbrot1968} (fBm) processes model the wandering of such laser beam, while the NTWS is a more general quantifier aimed to study any given dynamic system~\cite{paper:rosso2001}. Also, in a recent work we have analyzed the dynamic case: the laboratory-generated turbulence was set up to change in time~\cite{paper:zunino2005}. We have observed that these quantifiers are correlated, but at the time only a qualitative argument was given. Furthermore, each one of these quantifiers have been used separately to obtain information from biospeckle phenomenon~\cite{paper:passoni2004,paper:passoni2004a}.

The fBm is the only one family of processes which are self-similar, with stationary increments, and gaussian~\cite{book:taqqu94}. The normalized family of these gaussian processes, $B^H$, is the one with $B^H(0)=0$ almost surely, $\mathbb{E}[B^H(t)]=0$, and covariance
\begin{equation}
\E{B^H(t)B^H(s)}=\frac{1}{2}\left(\abs{t}^{2H}+\abs{s}^{2H}-\abs{t-s}^{2H}\right),
\label{eq:fbm-cov}\end{equation}
for $s,t\in\Re$. Here $\E{\,\cdot}$ refers to the average with gaussian probability density. The power exponent $H$ has a bounded range between $0$ and $1$. These processes exhibit memory, as can be observed from Eq.~\eqref{eq:fbm-cov}, for any Hurst paremeter but $H=1/2$. In this case successive Brownian motion increments are as likely to have the same sign as the opposite, and thus there is no correlation. Otherwise, it is the Brownian motion that splits the family of fBm processes in two. When $H>1/2$ the correlations of successive increments decay hyperbolically, and this sub-family of processes have long-memory. Besides, consecutive increments tend to have the same sign, these processes are \textit{persistent}. For $H<1/2$, the correlations of the increments also decay but exponentially, and this sub-family presents short-memory. But since consecutive increments are more likely to have opposite signs, it is said that these are \textit{anti-persitent}.

The Wavelet Analysis is one of the most useful tools when dealing with data samples. Thus, any signal can be descomposed by using a diadic discrete family $\{ 2^{j/2}\psi(2^j t -k) \}$---an  \textit{orthonormal} basis for $L^2(\mathbb{R})$---of translations and scaling functions based on a function $\psi$: the mother wavelet. This wavelet expansion has associated wavelet coefficients given by $C_j(k) = \langle \mathcal{S}, 2^{j/2}\psi(2^j \cdot -k) \rangle$.  Each resolution level $j$ has an associated energy given by $\mathcal{E}_j=\mathbb{E}\abs{C_j(k)}^2$. If the signal has stationary increments the coefficients are independent on $k$ and then the \textit{relative wavelet energy}, RWE, is
\begin{equation}
p_j=\frac{\mathcal{E}_j}{\mathcal{E}_\text{tot}},
\label{eq:rwe}\end{equation}
with $j\in\{-N,\dots,-1\}$, where $N = \log_2 M$ is the number of sample points, and $\mathcal{E}_\text{tot}=\sum_{j=-N}^{-1} \mathbb{E}\abs{C_j}^2$ is the total energy.																																																																																						       
Thus the NTWS is defined as (see \cite{paper:zunino2004} and references therein)
\begin{equation}
\label{eq:wav9}
  S_\text{WT}=  -\sum_{j=-N}^{-1}   p_j  \cdot \log_2 p_j/ S^\text{max},\quad \text{with} \quad S^\text{max}=\log_2 N.
\end{equation}

For a signal originated from a fBm the energy per resolution level can be calculated using the formalism introduced in Ref.~\cite{paper:perezzunino2004a}, see Appendix \ref{apx:a},
\begin{equation}
\mathbb{E}\abs{C_j(k)}^2 = 2\Gamma(2H+1)\sin(\pi H) 2^{-j(2H+1)}\int_0^{\infty} \frac{|\hat{\psi}|^2(\nu)}{\nu^{2H+1}}, 
\label{eq:re}
\end{equation}
for any mother wavelet election satisfying $\int_\Re \psi = 0$. From \eqref{eq:re} the relative wavelet energy for a finite data sample is
\begin{equation}
p_j = 2^{(j - 1)(1+ 2H)}\frac{1 - 2^{-(1+2H)}}{1- 2^{-N(1+2H)}},
\end{equation}
which becomes independent on wavelet basis. And so it does the normalized total wavelet entropy,
\begin{multline}
S_\text{WT}(N,H)= \frac{1}{\log_2 N} (1+ 2H)\left[\frac{1}{2^{1+2H}-1} - \frac{N}{2^{N(1+ 2H)}-1}\right] \\
- \frac{1}{\log_2 N}\log_2 \left[\frac{1-2^{-(1+2H)}}{1-2^{-N(1+2H)}}\right].
\label{eq:entrophy}
\end{multline}
As it was expected the entropy decreases when $H$ increases, with $H$ measuring the level of order of the signal.

\section{Simulations and tests}\label{sec:sim}

To test the functional relation between the Hurst exponent and NTWS we have simulated 50 fractional Brownian motion data samples~\cite{phd:coeurjolly2000} for each $H \in \{0.1, 0.2, 0.3, 0.4, 0.5, 0.6, 0.7, 0.8, 0.9\}$. Since we have examined data of $5000$ point in length in Ref.~\cite{paper:zunino2004}, these samples are set to the same length. For each set we estimate  $H$ and $S_\text{WT}$. Moreover, we employ an orthogonal cubic spline function as mother wavelet. Among several alternatives, the cubic spline function is symmetric and combine
in a suitable proportion smoothness with numerical advantages. It has become a recommendable tool for representing natural signals~\cite{paper:unser1999,paper:thevenaz2000}.

The Hurst parameter is estimated as usual: by plotting the logarithm of the estimated energy per resolution level $j$,
\begin{equation}
\widehat{\mathcal{E}}_j=\frac{1}{N_j}\sum^{N_j}_{k=1}\abs{C_j(k)}^2,
\end{equation}
with $N_j = 2^{-j}M$ the number of coefficients at resolution level $j$, versus $j$ and fitting a minimum square line. The slope of the line is the desired estimator.   

For NTWS we start dividing the signal into $N_T$ non-overlapping temporal windows with length $L$, $N_T = M / L$. The wavelet energy estimator at resolution level $j$ for the time window $i$ is given by
\begin{equation}
\label{eq:wav14}
\widehat{\mathcal{E}}^{(i)}_j={\frac{1}{N_j}}
\sum_{k=(i-1)\cdot L + 1}^{i \cdot L}C_j^2(k), 
\qquad
\text{with $i = 1, \cdots , N_T$} \ ,
\end{equation}
where $N_j$ represents the number of wavelet coefficients at resolution level $j$ corresponding to the time window $i$; while the total energy estimator in this time window will be $\widehat{\mathcal E}^{(i)}_{ \text{tot}} =\sum_{j < 0} \widehat{\mathcal{E}}^{(i)}_j$.

Hence, the time evolution estimators of RWE and NTWS  will be given by:
\begin{equation}
\label{eq:wav16}
\widehat{p}^{(i)}_j=\widehat{\mathcal{E}}^{(i)}_j / \widehat{\mathcal{E}}^{(i)}_{ \text{tot}},
\end{equation}
\begin{equation}
\label{eq:wav17}
\widehat{S}_\text{WT}{(i)}=  -\sum_{j<0} \widehat{p}^{(i)}_j \cdot \log_2 \widehat{p}^{(i)}_j/S^ \text{max}. 
\end{equation}

In order to obtain a quantifier for the whole time period under analysis
\cite{paper:rosso2001} the temporal average is evaluated. 
The temporal average of NTWS is given by
\begin{equation}
\label{eq:16}
\langle S_{ \text{WT}}\rangle = \frac{1}{N_T}\sum_{i=1}^{N_T} \widehat{S}_\text{WT}^{(i)},
\end{equation}
and for the wavelet energy at resolution level $j$
\begin{equation}
\label{eq:17}
\langle \mathcal{E}_j\rangle = \frac{1}{N_T}\sum_{i=1}^{N_T}\widehat{\mathcal{E}}^{(i)}_j;
\end{equation}
then the total wavelet energy temporal average is defined as $\langle \mathcal{E}_{ \text{tot}} \rangle=\sum_{j < 0} \langle \mathcal{E}_j\rangle$.
In consequence, a mean probability distribution $\{ q_j \}$
representative for the whole time interval (the complete signal) can be defined as
\begin{equation}
\label{eq:19}
q_j = \langle \mathcal{E}_j\rangle / \langle \mathcal{E}_{ \text{tot}} \rangle,
\end{equation}
with $\sum_{j} q_j = 1 $ and the corresponding mean NTWS as
\begin{equation}
\label{eq:20}
{\widetilde  S_{WT}}=-\sum_{j<0}  q_j \cdot \log_2 q_j/S^ \text{max}. 
\end{equation}

\begin{figure} 
\begin{center}
\psfrag{H}[][b]{$H$}
\psfrag{HE}{$\widehat{H}$}
\includegraphics[width=.8\textwidth]{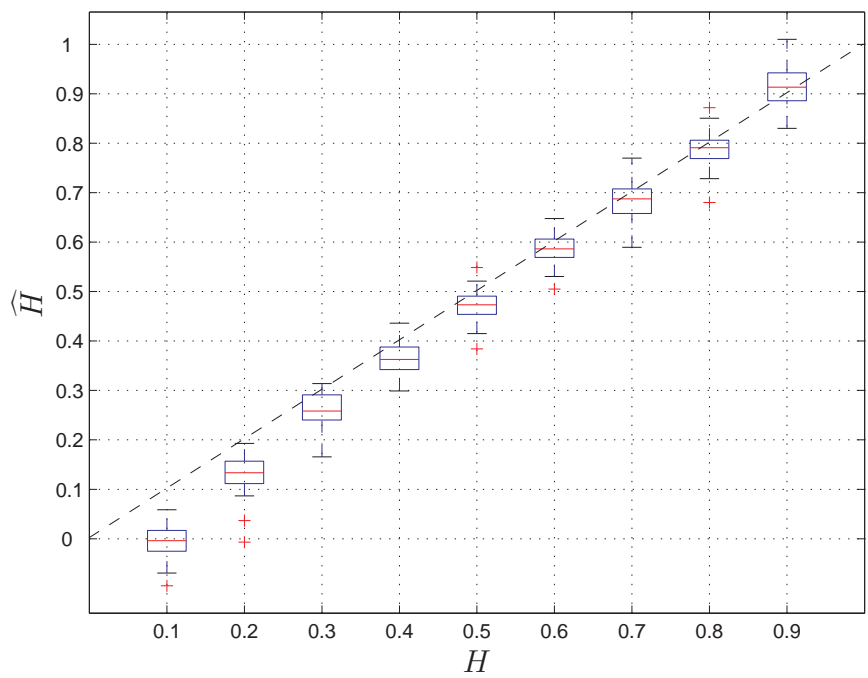}
\caption{The estimator for the Hurst parameter is plotted against the expected value, the dashed line is the identity.\label{figure:1}}
\end{center}
\end{figure}

\begin{figure} 
\begin{center}
\psfrag{NTWSb}[][t]{$\tilde{S}_\text{WT}$}
\psfrag{NTWS}[][t]{$\langle S_\text{WT}\rangle$}
\psfrag{H}[][b]{$H$}
     \subfigure{
	\includegraphics[width=.9\textwidth]{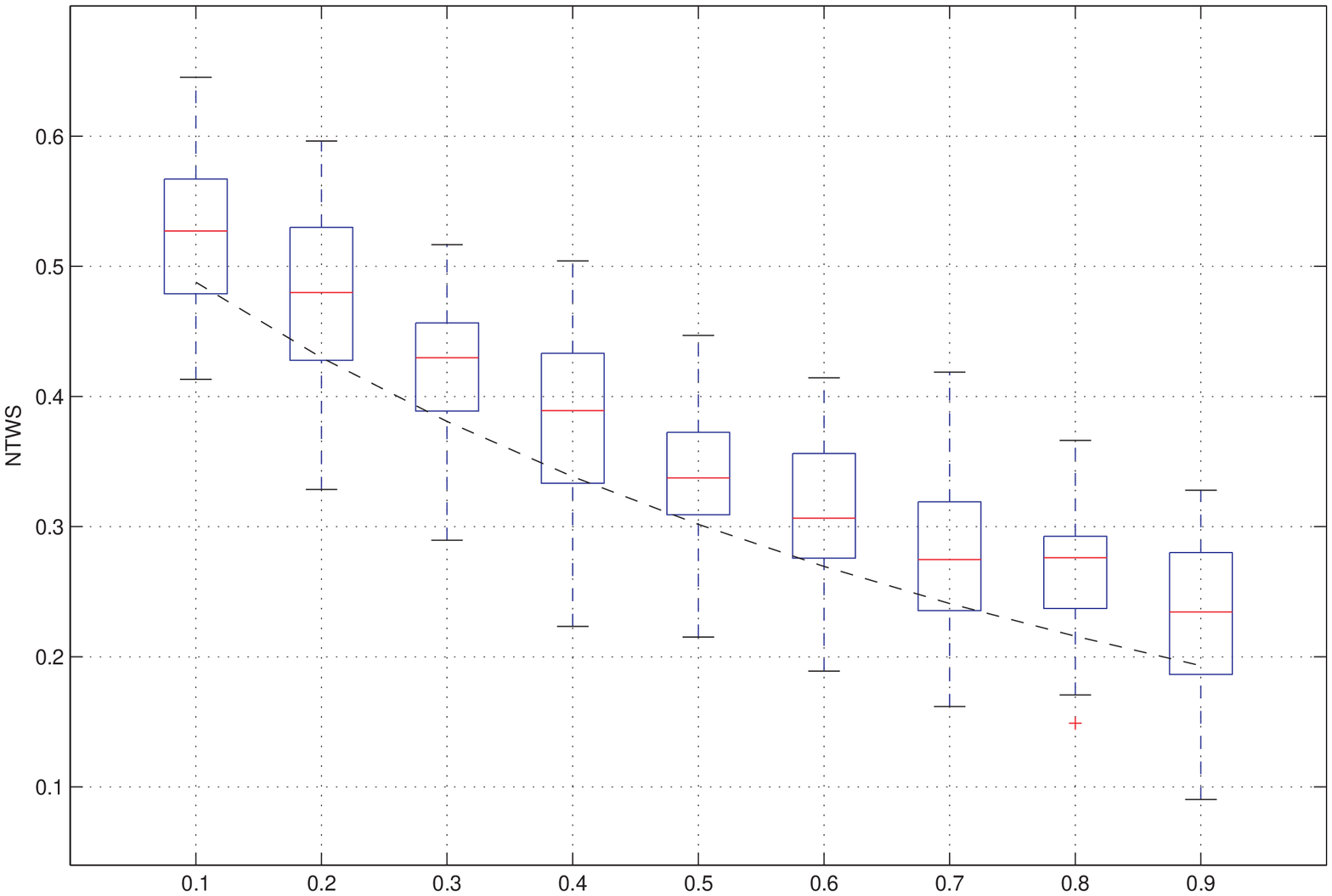}}
\vspace{.35cm}
     \subfigure{
         \includegraphics[width=.9\textwidth]{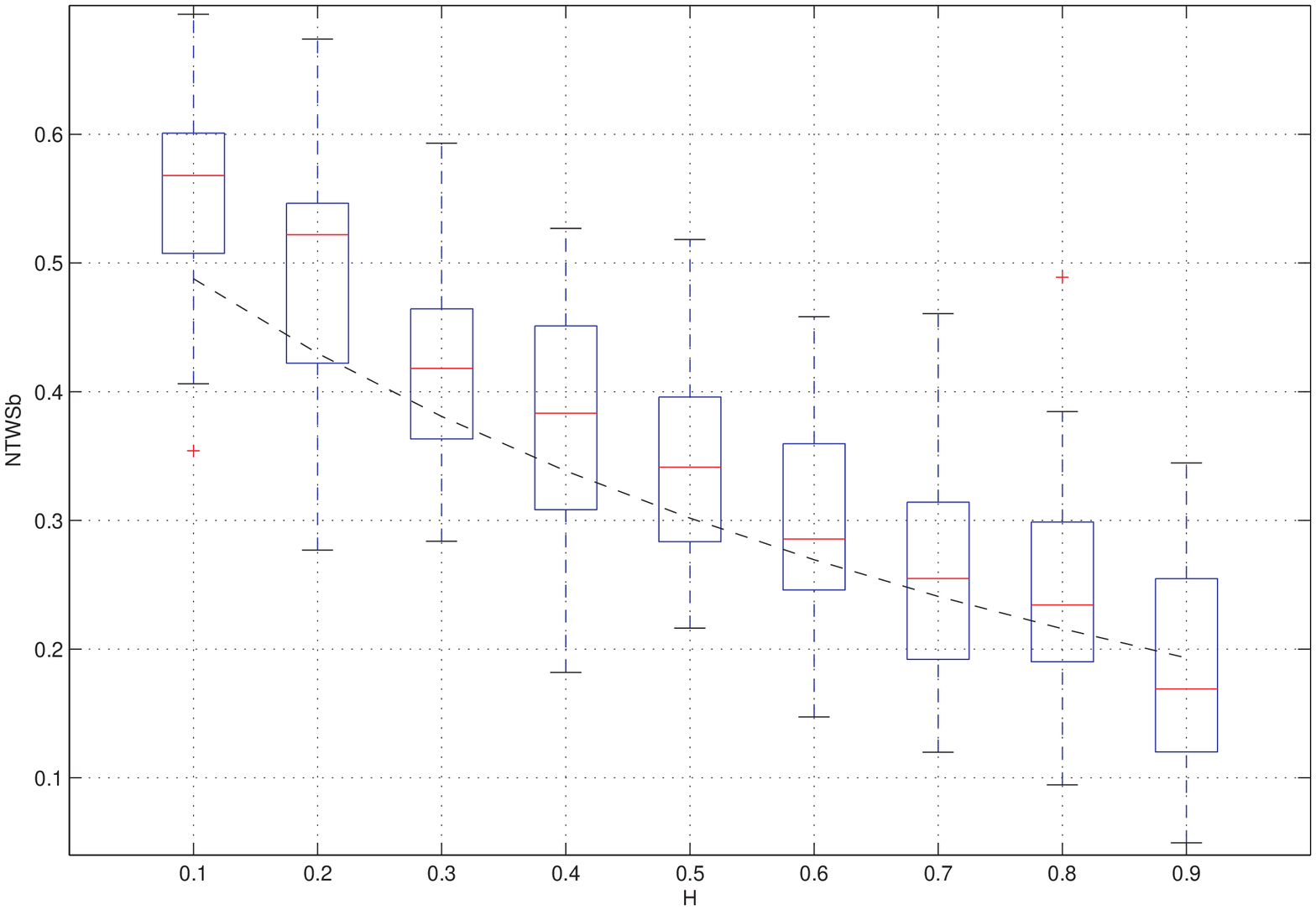}}
         
\caption{$\langle S_{ \text{WT}}\rangle$ (top) and $\tilde{S}_\text{WT}$ (bottom) are compared against $S_\text{WT}(12,H)$ (dashed line) as defined through eq.~\eqref{eq:entrophy}.\label{figure:2}}
\end{center}
\end{figure}

In Figure \ref{figure:1} we compare $H$ against its estimator. It has a good performance for $0.3<H<0.9$ and fails outside. Furthermore, the estimators fits better the larger values. Figure \ref{figure:2} represents the temporal average of NTWS, $\langle S_{ \text{WT}}\rangle$, and the mean NTWS, $\widetilde  S_{WT}$, estimated with our procedure and compared against the theoretical result in eq.~\eqref{eq:entrophy} with $N=12$. As usual, boxplots~\cite{book:tukey1977} show lower and upper lines at the lower quartile (25th percentile of the sample) and upper quartile (75th percentile of the sample) and the line in the middle of the box is the sample median. The whiskers are lines extending from each end of the box indicating the extent of the rest of the sample. Outliers are marked by plus signs. These points may be considered the result of a data entry error or a poor measurement.

\section{Conclusions}\label{sec:res}

For a fBm we have found, eq.~\eqref{eq:entrophy}, there is an inverse dependence: as $H$ grows the temporal average, $\langle S_{ \text{WT}}\rangle$, and mean NTWS, $\widetilde  S_{WT}$, diminishes. It is verified with the synthetic fBm data samples. This relation is logical, the spectrum has less high-frequency components as $H$ gets higher and all the energy is closer to the origin, and, if $H$ gets lower the energy contribution at high frequencies becomes relevant. Observe that the closer $\hat{H}$ is to the exact value, the better are the results for both estimators of the entropy.

From an analytical point of view both $H$ and $S_\text{WT}$ are equivalent. Although, the NTWS also contains information about the extension of the data set. Nevertheless, from a computational point of view the latter is independent on the scaling region, making the entropy less subjective. On the other hand the logarithm in the entropy definition introduces important errors, as we see in Figure \ref{figure:2}. To  narrow these it is necessary to increase the data samples. It should be stressed that extending the length of the data samples reduces the statistical error. 

\ack        
 
This work was partially supported by Consejo Nacional de Investigaciones Cient\'{\i}ficas y T\'ecnicas (CONICET, Argentina) and Pontificia Universidad Cat\'olica de Valpara\'iso (Project No. 123.774/2004, PUCV, Chile).

\section*{APPENDIX A}\label{apx:a}

Let us take as the signal $\mathcal{S}(t) = B^H(t,\omega)$, $\omega$ is fixed and represents one element of the statistic ensemble and it will be omitted hereafter. The wavelet coefficients are calculated using the orthonormal wavelet basis $\{2^{-j/2}\psi(2^{-j} \cdot - k)\}_{j,k\in\mathbbm{Z}}$, 
\begin{equation}
C^H_j(k) = \int_\Re 2^{-j/2}\psi(2^{-j} s - k) B^H(s)\d s 
= 2^{(1/2 + H)j }\int_\Re \psi(s) B^H(s + k) \d s,
\end{equation}
for the last step we used the self-similar property of the fBm; that is, $B^H(t)\overset{d}{=}c^H B^H(c^{-1}t)$. Since the fBm can be written, using the chaos expansion described in Ref.~\cite{paper:perezzunino2004a}, as
\begin{equation*}
B^H(t) = \sum^\infty_{n=1}\;\langle M_H\mathbbm{1}_{[0,t]},\xi_n\rangle\,\mathcal{H}_{\epsilon_n}(\omega),
\end{equation*}
where $\{\xi_n\}_{n\in\N}$ are the Hermite functions, and the operator $M_H$ is defined as follow~\cite{paper:elliot2003}
\begin{equation}
\widehat{M_H \phi}(\nu)= c_H \abs{\nu}^{1/2-H} \widehat{\phi}(\nu),
\label{eq:H-operator}\end{equation}
where the hat stands for the Fourier transform, $c_H^2= \Gamma(2H+1)\sin (\pi H)$, and $\phi$ is any function  in $L^2(\Re)$. Then, we introduce the following coefficients 
\begin{equation}
d^H_n(k) = \int_\Re \langle M_H\mathbbm{1}_{[0,s + k]},\xi_n\rangle \psi(s)\d s,
\end{equation}
to finally obtain:
\begin{equation}
C^H_j(k) = 2^{(1/2 + H)j }  \sum^\infty_{n=1}\;d^H_n(k)\,\mathcal{H}_{\epsilon_n}(\omega).
\label{eq:chaos}\end{equation}
The evaluation of the coefficients $d^H_n(k)$ is straightforward from their definition and eq.~\eqref{eq:H-operator}:
\begin{equation}
d^H_n(k) = -\frac{c_H}{i^n} \int_\Re \sgn\nu\, \abs{\nu}^{-(1/2 + H)} \Psi(\nu) \xi_n(\nu)\, e^{-ik\nu}\d \nu,
\label{eq:dcoeff}\end{equation}
where $\Psi(\nu)=\widehat{\psi}(\nu)$. 

The chaos expansion in eq.~\eqref{eq:chaos} corresponds to a Gaussian process, then for integers $j,k,j',k'$ the correlation is equal to~\cite{book:holden96}
\begin{equation*}
\E{C^H_j(k){C^{H}_{j'}}^\ast(k')} = \sum^\infty_{n=1} 2^{(1/2 + H)(j +j') } d^H_n(k)d^H_n(k').
\end{equation*}
Now, from eq.~\eqref{eq:dcoeff} and orthogonality of the Hermite functions the above equation is rewritten in the following way
\begin{equation}
\E{C^H_j(k){C^{H}_{j'}}^\ast(k')} =c^2_H  2^{(1/2 + H)(j +j') } \int_\Re \abs{\nu}^{-(1 + 2H)} \abs{\Psi(\nu)}^2 e^{-i(k - k')\nu}\d \nu,
\label{eq:final}\end{equation}
which is the usual expresion found in many works, see \cite{paper:flandrin1992} and references therein. The integral has convergence problems near the origin. These are resolved chosing a mother wavelet $\phi$ with $K$ null moments. That is,
\begin{equation*}
\int_\Re t^k \phi(t)\,\d t =0,
\end{equation*}
for $k = 0,\cdots, K-1$. Therefore, $\abs{\Psi(\nu)}^2 = a_1 \abs{\nu}^{2K} + a_2 \abs{\nu}^{2K + 1} + \mathit{o}(\abs{\nu}^{2K + 1})$. When $k$ and $k'$ are far apart, i.~e., $m = k - k'\rightarrow \infty$, the integral in eq.~\eqref{eq:final} is dominated by the contribution of frequencies in the interval $[0,1]$, thus giving
\begin{multline}
\E{C^H_j(k){C^{H}_{j'}}^\ast(k')}  \approx 2a_1 c^2_H  2^{(1/2 + H)(j +j') } \int^1_0 \nu^{2K -2H -1}\cos(m\nu) \d\nu\\
= 2a_1 c^2_H  2^{(1/2 + H)(j +j') } \Gamma(2K-2H) \cos(\pi (K -H)) m^{2K - 2H} \\
+ \mathcal{O}(m^{2K - 2H-1}),
\end{multline}
for $K>H$. The coefficients of a wavelet expansion are highly correlated. But, for $j=j'$ and $k =k'$, 
\begin{align}
\mathbbm{E}\abs{C^H_j(k)}^2 &=c^2_H  2^{(1 + 2H)j  } \int_\Re \abs{\nu}^{-(1 + 2H)} \abs{\Psi(\nu)}^2\d \nu\nonumber\\
&= 2 \Gamma(2H+1)\sin (\pi H) 2^{(1 + 2H)j  } \int_0^\infty \nu^{-(1 + 2H)} \abs{\Psi(\nu)}^2\d \nu,
\end{align}
we recover the mean energy by resolution level $j$. Therefore, the RWE is obtained replacing the above  into eq.~\eqref{eq:rwe}:
\begin{equation}
p_j = \frac{2^{j(1+2H)}}{\sum^{-1}_{j=-N} 2^{j(1+2H)}} = \left[\frac{1-2^{-(1+2H)}}{1-2^{-N(1+2H)}}\right]2^{(j+1)(1+2H)}.
\end{equation}
where the last equation comes from the evaluation of the geometric series corresponding to the total energy. Its logarithm (base 2) is simply $\log_2 p_j = (1+2H)(j+1) + \log_2[1-2^{-(1+2H)}/1-2^{-N(1+2H)}]$. Finally, using these results in the definition of NTWS, it is
\begin{multline}
S_\text{WT}(N,H)=  \left[\frac{1-2^{-(1+2H)}}{1-2^{-N(1+2H)}}\right] (1+2H) \sum^{-1}_{j=-N} (j+1) 2^{(j+1)(1+2H)} \\
+ \left[\frac{1-2^{-(1+2H)}}{1-2^{-N(1+2H)}}\right]\log_2\left[\frac{1-2^{-(1+2H)}}{1-2^{-N(1+2H)}}\right] \sum^{-1}_{j=-N} 2^{(j+1)(1+2H)},
\end{multline}
then handling the geometric sums carefully we obtain eq.~\eqref{eq:entrophy}.

\bibliography{references}

\begin{thebibliography}{10}
\expandafter\ifx\csname url\endcsname\relax
  \def\url#1{\texttt{#1}}\fi
\expandafter\ifx\csname urlprefix\endcsname\relax\def\urlprefix{URL }\fi

\bibitem{paper:zunino2004}
L.~Zunino, D.~G. P{\'e}rez, M.~Garavaglia, O.~A. Rosso, Characterization of
  laser propagation through turbulent media by quantifiers based on the wavelet
  transform, Fractals 12~(2) (2004) 223--233.

\bibitem{paper:mandelbrot1968}
B.~B. Mandelbrot, J.~W.~V. Ness, Fractional {B}rownian motions, fractional
  noises and applications, SIAM Rev. 4 (1968) 422--437.

\bibitem{paper:rosso2001}
O.~A. Rosso, S.~Blanco, J.~Yordanova, V.~Kolev, A.~Figliola, M.~{Sch\"urmann},
  E.~{Ba\c{s}ar}, Wavelet entropy: a new tool for analysis of short duration
  brain electrical signals, J. Neuroscience Method 105 (2001) 65--75.

\bibitem{paper:zunino2005}
L.~Zunino, D.~G. P{\'e}rez, M.~Garavaglia, O.~A. Rosso, Characterization of
  laser propagation through turbulent media by quantifiers based on the wavelet
  transform: dynamic study, submitted to Physica A (January 2005).

\bibitem{paper:passoni2004}
I.~Passoni, H.~Rabal, C.~M. Arizmendi, {Characterizing dynamic speckle time
  series with the Hurst coefficient concept}, Fractals 12~(3) (2004) 319--329.

\bibitem{paper:passoni2004a}
I.~Passoni, A.~{Dai Pra}, H.~Rabal, M.~Trivi, R.~Arizaga, Dynamic speckle
  processing using wavelets based entropy, Optics Communications (in press,
  2005).

\bibitem{book:taqqu94}
G.~Samorodnitsky, M.~S. Taqqu, Stable non-Gaussian random processes, Stochastic
  Modeling, Chapman {\&} Hall, London, U.K., 1994.

\bibitem{paper:perezzunino2004a}
D.~G. P{\'e}rez, L.~Zunino, M.~Garavaglia, Modeling the turbulent wave-front
  phase as a fractional brownian motion: a new approach, J. Opt. Soc. Am. A
  21~(10) (2004) 1962--1969.

\bibitem{phd:coeurjolly2000}
J.-F. Coeurjolly, Statistical inference for fractional and multifractional
  {Brownian} motions, Ph.D. thesis, Laboratoire de Mod\'elisation et Calcul -
  Institut d'Informatique el Math\'ematiques Appliqu\'ees de Grenoble (2000).
\newline\urlprefix\url{http://bibliotheque.imag.fr/publications/theses/2000}

\bibitem{paper:unser1999}
M.~Unser, Spline: a perfect fit for signal and image processing, IEEE Signal
  Processing Magazine 16 (1999) 22--38.

\bibitem{paper:thevenaz2000}
P.~Th\'evenaz, T.~Blu, M.~Unser, Interpolation revisited, IEEE Trans. on
  Medical Imaging 19~(7) (2000) 739--758.

\bibitem{book:tukey1977}
{J. W. Tukey}, Exploratory Data Analysis, Addison-Wesley, 1977.

\bibitem{paper:elliot2003}
R.~J. Elliott, J.~van~der Hoek, A general fractional white noise theory and
  applications to finance, Mathematical Finance 13 (2003) 301--330.

\bibitem{book:holden96}
H.~Holden, B.~{\O}ksendal, J.~Ub{\o}e, T.~Zhang, Stochastic partial
  differential equations: A modeling, white noise functional approach, in:
  Probability and Its Applications, Probability and Its Applications,
  Birkh{\"a}user, 1996.

\bibitem{paper:flandrin1992}
P.~Flandrin, Wavelet analysis and synthesis of fractional {Brownian} motion,
  IEEE Trans. Inform. Theory IT-38~(2) (1992) 910--917.

\end{thebibliography}
\bibliographystyle{elsart-num}

\newpage

\end{document}